\documentclass[preprint,showpacs,superscriptaddress,preprintnumbers,amsmath,amssymb]{revtex4}

\usepackage{graphicx}
\usepackage{dcolumn}
\usepackage{bm}
\usepackage{float}
\usepackage{placeins}
\usepackage{lscape}
\usepackage{epstopdf}
\usepackage{color}
\usepackage{ulem}
\usepackage{multirow}

\usepackage[font = normalsize]{caption} 

\begin{document}

\title {Supercoupling between heavy-hole and light-hole states in self-assembled quantum dots }
\author {Jun-Wei Luo}
\email{jwluo@semi.ac.cn}
\address {State key laboratory of superlattices and microstructures, Institute of Semiconductors, Chinese Academy of Sciences, Beijing 100083, China}
\address {Synergetic Innovation Center of Quantum Information and Quantum Physics, University of Science and Technology of China, Hefei, Anhui 230026, China}
\author{Gabriel Bester}
\affiliation{Max-Planck-Institut f\"{u}r Festk\"{o}rperforschung, Heisenbergstrasse 1, D-70569 Stuttgart, Germany.}
\author{Alex Zunger}
\address{University of Colorado, Boulder, Colorado 80309, USA}

\date{\today }

\begin{abstract}
Spintronics, quantum computing and quantum communication science utilizing cubic semiconductors rely largely on the properties of the hole states, composed of light and heavy hole wavefunction components. The admixture of light-hole (LH) into ground hole state predominately by the heavy hole (HH) would induce unique features of LH in optical transitions, spin relaxation, and spin polarization. We point to an unexpected source of HH-LH mixing in quantum dots, arguing that in contrast with current models the mixing does not reflect the strain between the dot and its matrix and does not scale inversely with the energy splitting between the bulk HH and LH states. Instead, we show via atomistic pseudopotential calculations on a range of strained and unstrained dots of different symmetries that the HH-LH mixing is enabled by the presence in the QD of a dense ladder of intermediate states between the HH and LH states which amplifies and propagates this interaction and leads to "supercoupling" (analogous to super-exchange in magnetism). This explains a number of outstanding puzzles regarding the surprising large coupling seen in unstrained QD (GaAs/AlAs) of ideal shapes and the surprising fact that in strained QD (InAs/GaAs) the coupling is very strong despite the fact that the 12-fold increase in bulk HH-LH splitting overrides the ~4 fold enhancement of the coupling matrix element by strain in comparison with unstrained GaAs QDs.
\end{abstract}

\maketitle

The reduced symmetry in low-dimensional nanostructures with respect to 3D bulk crystals offers the possibility of quantum mixing between the "heavy-hole" (HH) and "light-hole" (LH)  components  of the bulk $\Gamma_{8v}$ valence band  as its dimensionality  is reduced from 3D bulk  to 0D quantum dots (QDs). Such a HH-LH quantum mixing is expected to have profound  effects  on  properties of QDs \cite{huo14}, including (i)   tuning of   the  excitonic fine-structure  splitting   \cite{bennett10,ghali12,luo12} which controls  the fidelity  of  entangled  photon  pairs,  (ii)  providing an efficient channel for  the spin  decoherence   \cite{Chekhovich13,Warburton13, Testelin09,Eble09, fischer10, luo10}, (iii) creating   a  polarization  anisotropy of  light emission which  is  important  for  quantum  information  schemes    \cite{koudinov04, kowalik08, belhadj10, tonin12,liao12}, and (iv)  giving an additional efficient mechanism for  the optical initialization of hole  spin  qubit   \cite{xu07, Fras13}.  Despite  the    important  role of HH-LH mixing   in  QDs,  the understanding of the basic   experimental  observations  pertaining  to  such  mixing  remains unclear.

\begin{figure}[H]
\centering
\includegraphics[width=0.6\textwidth]{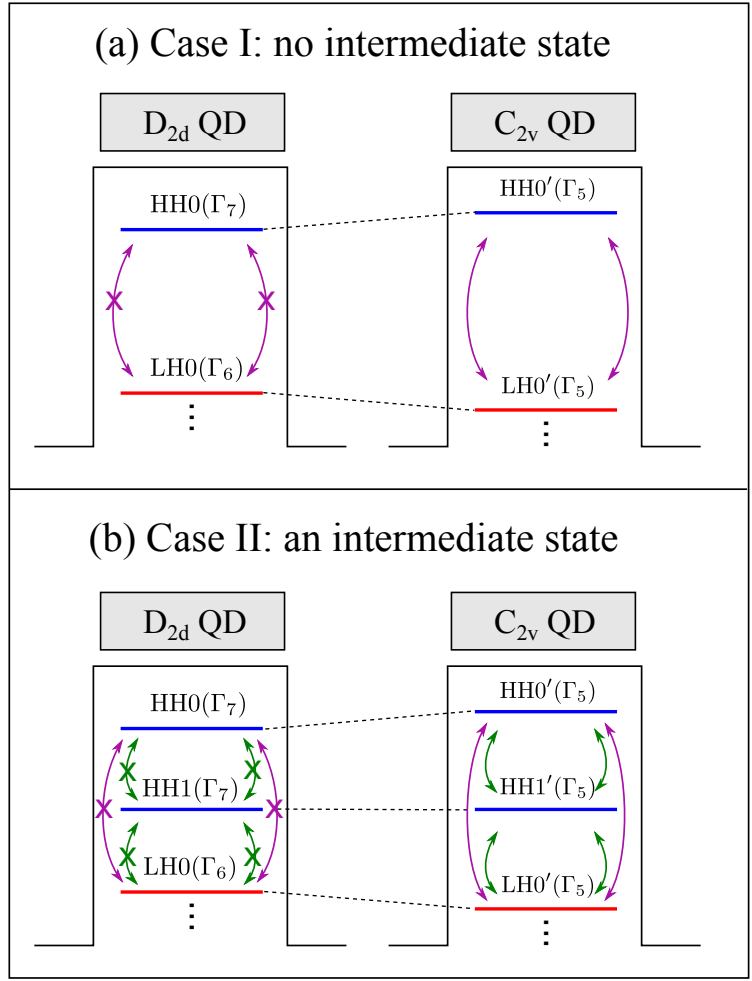}
\caption{\label{fig:schematic} Schematic electronic structure of valence bands in SAQDs. (a) Case I is like in common  perception there is no intermediate states lying between HH0 and LH0 states. (b) Case II is usually in strained QDs where strained induced large HH-LH splitting accommodate the existence of intermediate states lying between HH0 and LH0 states. In $D_{2d}$ symmetry the coupling between HH and LH states are forbidden and in $C_{2v}$ such coupling is allowed.}
\end{figure}

\begin{figure}[H]
\centering
\includegraphics[width=0.8\textwidth]{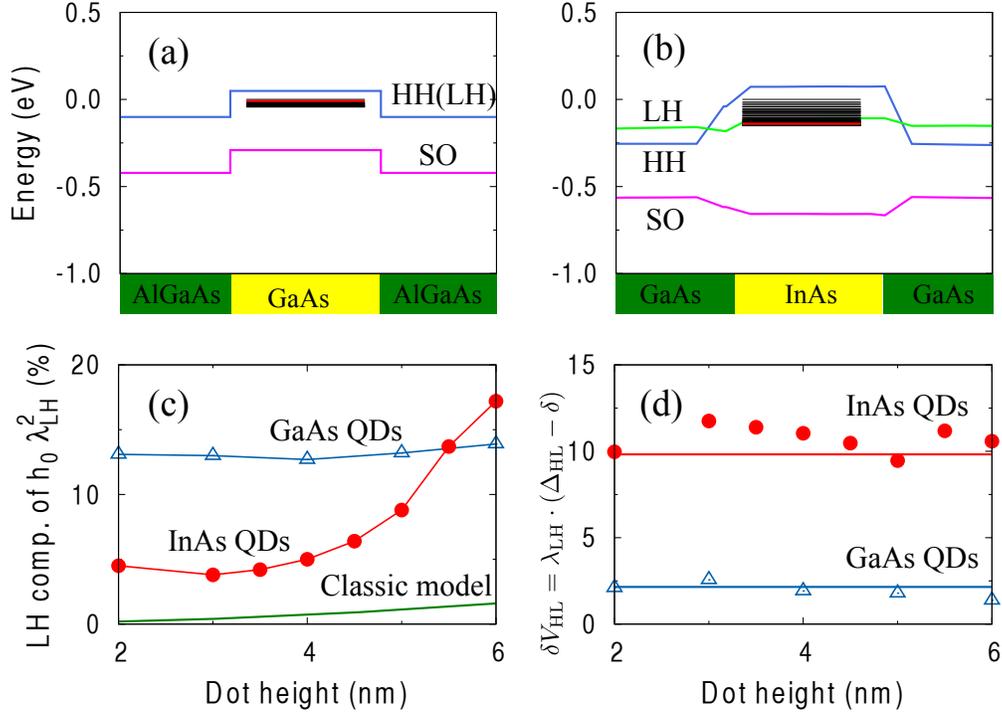}
\caption{\label{fig:BandOffsets} Band alignment and energy level structure of strained and unstrained QDs obtained from direct atomistic calculations: (a)  unstrained GaAs/AlGaAs QD and (b) strained InAs/GaAs QD.   Short lines are for QD electron and hole levels and red one for LH-like ground state. Bold lines represent energy alignment of HH, LH, and SO through SAQDs centre, (c) QD-height dependence of $\lambda^2_{\textrm{LH}}$ of  GaAs/Al(Ga)As QDs and  InAs/GaAs QDs (with base size of 25 nm) as well as results of classic description for GaAs QDs \cite{fischer10}.  (b) QD-height dependence of $\delta V_{\textrm{HL}}$  directly extracted from Eq. ~\ref{eq4-1} (unstrained QDs) or Eq. ~\ref{eq:super} (strained QDs). Two bold lines indicate fitted common $\delta V_{\textrm{HL}} $ values of two corresponding QD classes.}
\end{figure}

It is a common perception that  the  admixture  of  LH component  into  the ground hole with dominated HH component   in  a  QD  would scale as $\lambda^2_{\textrm{LH}}=(\delta V_{\text{HL}}/\Delta_{\textrm{HL}})^2$, where $\delta V_{\text{HL}}$ is the {\it coupling matrix element}  between unperturbed HH and LH ground states, and  $\Delta_{\textrm{HL}}$ (or termed  {\it HH-LH splitting}) is the energy separation between them.  To have finite HH-LH mixing the symmetry-controlled  coupling matrix element $\delta V_{\text{HL}}$ must be non-vanishing.  Self-assembled QDs (SAQDs) are usually found to be lens-shape \cite{kowalik08,koudinov04}  or Gaussian- shape \cite{belhadj10} and improperly recognized,  in continuum point of view, as to be $D_{2d}$ symmetry  \cite{belhadj10,kowalik08,koudinov04}, a group in which HH and LH states belong to  different symmetry representations ($\Gamma_{7v}$ and $\Gamma_{6v}$, respectively) and  consequently, just as in the parent bulk compounds, these states can not mix as shown in Fig.~\ref{fig:schematic}(a).  Such expectations  lead  to the attribution  of  experimentally  observed  HH-LH  mixing  in strained self-assembled In(Ga)As/GaAs \cite{kowalik08} and CdTe/ZnTe \cite{koudinov04} QDs  to strain-induced symmetry lowering  below  $D_{2d}$ \cite{kowalik08, koudinov04}, or in unstrained GaAs/AlGaAs QDs to  a  presumed  shape-distortion of the disk-like symmetry, e.g,  through elongations of the QDs \cite{belhadj10, note1}.   The fact, in the atomistic point of view, is that  the  symmetry  of  even unstrained and ideally  shaped  (circular based  lens-, cone-, or Gaussian-shape)  SAQDs made of zinc-blende (ZB) semiconductors is already lower to $C_{2v}$, a group in which all QD states, including both HH and LH states, belong to its sole symmetry representation $\Gamma_{5}$ and  consequently,  they are allowed to mix each other [Fig.~\ref{fig:schematic}(a)].  Because SAQDs with curved upper interface are distinct from bulk ZB crystals or $D_{2d}$ symmetric (001) quantum wells, where the (110) plane can be transformed to the (1$\bar1$0) plane by an $S_4$ symmetry operations  \cite{Koster63} (90$^\circ$ rotation followed by a  reflection across a mirror plane perpendicular to both (110) and (1$\bar1$0) planes),  the equivalence of the crystal fields in (110) and (1$\bar1$0) planes through QD center is lifted, leading ideally shaped QDs to $C_{2v}$. The built-in strain indeed enhances the HH-LH coupling matrix element $\delta V_{\text{HL}}$,  it also would significantly increase the HH-LH splitting $\Delta_{\textrm{HL}}$, as evident from atomistic pseudopotential calculations on In(Ga)As/GaAs QDs \cite{wei94, Bir-Pikus} showing that  the   HH  and  LH  bands, which are  degenerate in bulk InAs,  are split by  as  much  as  $\Delta_{\text{HL}}=120$ meV as shown in  Fig.~\ref{fig:BandOffsets}. It suggests  that the strain enhanced numerator $\delta V_{\text{HL}}$ would be overridden by strain magnified denominator $\Delta_{\text{HL}}$ leading to {\it diminishing} HH-LH mixing.  Consequently, the LH mixing was neglected altogether in the early days of QD physics \cite{Jacak98}.  Moreover, the large HH-LH splitting lifted by built-in strain unavoidably causes the existence of the HH-like intermediate states lying between HH- and LH-ground states.  The possible influence of these intermediate states on HH-LH mixing has so far never been discussed due to two-level models were adopted exclusively  to describe it in literature. Here we address the fundamental mechanism leading to sizeable HH-LH mixing observed in strained SAQDs and the effect of intermediate HH-like states on HH-LH mixing [Fig.~\ref{fig:schematic}(b)].

Our strategy for gaining access to the physics determining the HH-LH mixing in QDs is to first calculate  the wave functions in a large range of shapes, compositions and strain of (Ga,In)As dots in a (Al,Ga)As matrix, using the high precise  atomistic  pseudopotential theory, free from any specific model assumptions on the nature of the HH-LH mixing. We  address  the  QD  problem  by  solving  the  multi-million  atom QD as if it were a giant molecule with discrete atoms that are located at specific  positions, each carrying its own (screened) pseudopotential. The total potential appearing in the Schr\"odinger equation is a superposition of atomic pseudopotentials (including spin-orbit coupling) located at relaxed (possibly strained) positions. This description  forces  upon us   the  correct  atomically-resolved symmetry,  thereby  including  automatically  effects  of  shape,  strain,   alloy  fluctuations, and  wave  function  mixing. The calculated eigenvalues are shown in Fig.~\ref{fig:BandOffsets}. Since we use  explicitly  the  microscopic  potential  of  the  QD  system  under  consideration,  we are  free from  the  need  to  pre-judge at  the  outset  which    3D  bands  will  couple  in  0D; this determination is done instead via analysis after the direct diagonalization is completed. The HH-LH mixing, as well as other inter-band coupling, is already  present in QD states from the direct atomistic calculation. In order to retrieve HH and LH components we  project the QD wave functions onto a basis of  bulk bands, such as HH = $|3/2,\pm3/2\rangle$, LH = $|3/2,\pm1/2\rangle$, SO = $|1/2,\pm1/2\rangle$, and  conduction bands at the $\Gamma$-point \cite{bester03}.   We now obtain (i) HH-LH splitting $\Delta_{\text{HL}}$ and (ii) the magnitude of HH-LH mixing $\lambda^2_{\textrm{LH}}$.

The QD hole states of direct atomistic  calculations are next map onto classic description of HH-LH mixing, backing out   the coupling matrix element $\delta V_{\text{HL}}$ for a QD class (or an ensemble), and establishing  the  various  physical  factors  contributing  to  such coupling.  In the classic (but  much  simplified)  descriptions in terms of perturbation theory  \cite{bennett10,ghali12,luo12, fischer10,koudinov04, kowalik08, belhadj10,liao12,Fras13,Chekhovich13,Warburton13,huo14}, the model Hamiltonian $H_{C_{2v}}$ of a $C_{2v}$ symmetry QD is  divided into two parts: $H_{C_{2v}}=H_0+\delta V_{C_{2v}}$, where $H_0$ is the bare Hamiltonian for QD overestimated to be $D_{2d}$, and has eigenstates of  unperturbed HH and LH states: $|\Psi^0_{\textrm{HH}n}\rangle$ and $|\Psi^0_{\textrm{LH}n}\rangle$ ($n=0,1,2, ...$). The symmetry-lowering perturbation potential $\delta V_{C_{2v}}$ introduces inter-band coupling. The QD  state   $|\Psi'_{\textrm{HH}0}\rangle$ (the  QD  ground  hole  state  $h_0$) can then be written as:
\begin{equation}
\label{eq4-2}
|\Psi'_{\textrm{HH}0}\rangle = |\Psi^0_{\textrm{HH}0}\rangle +\sum_m \frac{\langle \Psi^0_{\textrm{LH}m}|\delta V_{C_{2v}}|\Psi^0_{\textrm{HH}0}\rangle}{E^0_{\textrm{HH}0}-E^0_{\textrm{LH}m}}|\Psi^0_{\textrm{LH}m}\rangle.
\end{equation}
In the classic description of QDs, the HH0 state is always presumed to be adjacent by LH0 in energy, a schematic case shown in  Fig.~\ref{fig:schematic}(a).  Due to larger energy separation of LH excited states from HH0,  their coupling to HH0 could be neglected in comparison to LH0. Therefore, $h_0$ is approximated as \cite{koudinov04, fischer10, luo10}: 
\begin{equation}
\label{eq4-0}
|\Psi_{h_0}\rangle=| \Psi^0_{\textrm{HH0}}\rangle + \lambda_{\textrm{LH}}|\Psi^0_{\textrm{LH0}}\rangle, 
\end{equation}
where  $\lambda_{\textrm{LH}}$ is a mixing coefficient  given by 
\begin{equation}
\label{eq4-1}
\lambda_{\textrm{LH}} = \frac{\langle \Psi^0_{\textrm{LH0}} |\delta V_{C_{2v}}| \Psi^0_{\textrm{HH0}}\rangle}{E^0_{\textrm{HH0}}-E^0_{\textrm{LH0}}} = \frac{\delta V_{\textrm{HL}}}{\Delta_{\textrm{HL}}},
\end{equation}
and $\Delta_{\textrm{HL}}$ is the energy separation between unperturbed HH0 and LH0 states.  By inserting the values of $\lambda_{\textrm{LH}}$ and $\Delta_{\text{HL}}$, retrieved from direct  atomistic  calculations, into  Eq.~(\ref{eq4-1})  the effective coupling matrix element $\delta V_{\text{HL}}$ is obtained for individual QDs. This  strategy,  of  first  securing  the  least  approximated  description,  followed  by  its  analytic  dissection  into  the simplified  language of  Eqs.~(\ref{eq4-0}) and ~(\ref{eq4-1}),  allows  us  to  get properly  coupling  matrix  element  $\delta V_{\text{HL}}$ while providing  the  necessary  communication  with  the  classic  literature  based  on  constructing  the  full answer  from the  simplified  description itself.

\begin{figure}[H]
\centering
\includegraphics[width=0.5\textwidth]{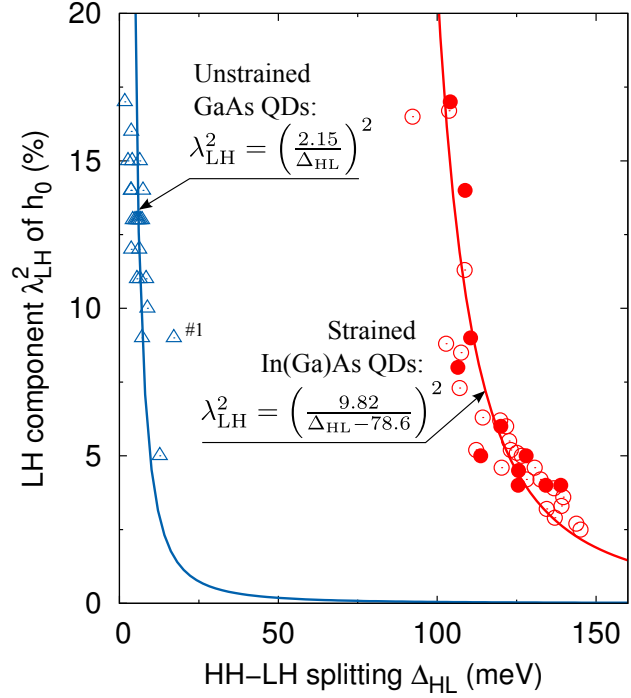}
\caption{\label{fig:Mixing} HH-LH  mixing strength $\lambda^2_{\textrm{LH}}$ of the QD's ground hole states $h_0$ obtained from direct atomistic calculations. {\bf a.}   $\lambda^2_{\textrm{LH}}$ {\it vs}  $\Delta_{\textrm{HL}}$ for both strained  In(Ga)As/GaAs (represented by red dots and circles) and unstrained GaAs/Al(Ga)As QDs (by blue triangles) with varying QD shape, size and composition. All GaAs/Al(Ga)As QDs have their $\lambda^2_{\textrm{LH}}$ {\it vs} $\Delta_{\text{HL}}$ values close to a common curve described by Eq. ~(\ref{eq4-1})  with $\delta V_{\textrm{HL}} = 2.15$ meV, whereas all the data of strained In(Ga)As/GaAs QDs are well described by a common curve given by Eq. ~(\ref{eq:super})  with $\delta V_{\textrm{HL}} = 9.82$ meV. }
\end{figure}

We consider two types of SAQDs \cite{note2}: (i)  {\it unstrained}  Gaussian-shaped  GaAs/Al(Ga)As  QDs \cite{belhadj10}  and (ii) {\it strained} lens-shaped In(Ga)As/GaAs  QDs \cite{kowalik08,koudinov04}, with varying QD height, base size, and compositions, both belonging to nominal $C_{2v}$ symmetry (here, "nominal symmetry" refers to QD symmetry excluding alloying effect).  Fig.~\ref{fig:Mixing}  shows as blue triangles $\lambda_{\textrm{LH}}$ {\it vs} $\Delta_{\text{HL}}$ for 24 {\it unstrained} GaAs/Al(Ga)As  QDs, whereas  the results  of 37 {\it strained}  In(Ga)As/GaAs  QDs  are  represented  by  red  dots or circles.  In  the class of unstrained GaAs/AlAs QDs, despite their different shapes, sizes, and compositions, the $\lambda_{\textrm{LH}}$ values of  all such QDs fall close to a common curve given by Eq.~(\ref{eq4-1}) with a {\it common} coupling matrix element $\delta V_{\textrm{HL}} = 2.15$ meV (except QD \#1  for  which  an explanation  will  follow).  All data points of strained InAs QDs shown in Fig.~\ref{fig:Mixing}, however, exhibit a blue shift by an energy of $\delta=78.6$  meV with respect to the class of unstrained GaAs QDs and fall close to another curve:
\begin{equation}
\label{eq:super}
\lambda_{\textrm{LH}} = \frac{\delta V_{\textrm{HL}}}{(\Delta_{\textrm{HL}}-\delta)}, \quad 
\end{equation}
with $\delta V_{\textrm{HL}} = 9.82$ meV.  It is interesting to notice the existence of a common value of $\delta V_{\textrm{HL}} $ for an entire class of QDs, which suggests that QD sizes, shape distortion, and alloy compositions do not influence remarkably on the coupling  matrix $\delta V_{\textrm{HL}}$, in contrast with earlier expectations \cite{fischer10}.  By inserting $\lambda_{\textrm{LH}}$ and $\delta V_{\textrm{HL}}$ of individual QDs into Eqs. (2) and (3), respectively, for unstrained and strained QDs, $\delta V_{\textrm{HL}} $ of individual QDs is ready to earn. Fig.~\ref{fig:BandOffsets} (d) shows calculated  $\delta V_{\textrm{HL}}$ values of individual unstrained GaAs QDs and of strained InAs QDs with a fixed base size and varying the QD height from 2 to 6 nm, as well as fitted common $\delta V_{\textrm{HL}}$ of respective QD classes. It indeed exhibits that $\delta V_{\textrm{HL}}$ of unstrained GaAs QDs lies on a line of 2.15 meV and of strained InAs QDs around another common value of 9.82 meV,  in sharp contrast to classic model-Hamiltonian considerations \cite{fischer10}, where all possible coupling terms entering the model are expected to be strongly dependent on  QD-height.  This confirms the existence of a common coupling matrix $\delta V_{\textrm{HL}}$ for all QDs within a class.  Fig.~\ref{fig:BandOffsets} (c) shows that  in the case of strained InAs QDs, $\lambda_{\textrm{LH}}$ increases as QD height meanwhile  individual QDs of $\delta V_{\textrm{HL}} $ are around the value of 9.82 meV, implying reduced strain by increasing QD height mostly reduces $\Delta_{\textrm{HL}}$ but not $\delta V_{\textrm{HL}} $.  Whereas, in the case of  unstrained GaAs QDs $\Delta_{\textrm{HL}}$ (and $\lambda_{\textrm{LH}}$) is QD-height independent,  illustrating the negligible effect of quantum confinement on $\Delta_{\textrm{HL}}$. It is again distinct from common thought.

\paragraph*{Supercoupling between HH and LH mediated by intermediate  states  in strained QDs.} The  reduction  of  the  HH-LH  splitting  in Eq.~(\ref{eq:super})  implies a  novel  effect that will  be  called  "supercoupling",  {\it whereby a highly dense manifold of HH-like QD states lying in energy between the primary HH and LH ground states mediates the HH-LH coupling and significantly enhances the mixing} by reducing  the  energy-denominator  $\Delta_{\text{HL}}$ to  a smaller  effective value  $\Delta_{\text{eff}}=\Delta_{\text{HL}}-\delta$. Fig.~\ref{fig:BandOffsets} demonstrates the existence of a dense manifold of such intermediate  hole states obtained from an atomistic calculation of a realistic lens-shaped  InAs QD  strained in  a  coherent GaAs matrix. We see in Fig.~\ref{fig:BandOffsets}(b), but not in Fig.~\ref{fig:BandOffsets}(a) where LH-like state is just adjacent to the HH-like ground state, a dense manifold of states, derived predominantly from bulk HH band, lying between HH- and LH-like ground states. These HH-like intermediate  states are the agents of the supercoupling between HH and LH in strained  QDs. This  supercoupling  effect ($\delta=\Delta_{\text{HL}}-\Delta_{\text{eff}}$) is  identical to all QDs within  an  entire  class (or ensemble), e.g. $\delta=78.6$ meV for strained $C_{2v}$ In(Ga)As/GaAs  QDs with  varying sizes,  shape distortions  and  alloy compositions, whenever the  fluctuation in the number of intermediate states is small ($\sim1$\%). It should be noted that for strained In(Ga)As/GaAs QDs, the curve is only  fitted to 11 QDs indicated by red dots (presented in elsewhere \cite{Luobookchapter}), and remaining 26 QDs indicated by red circles are calculated after fitting. The right description of the latter QDs by fitted curve manifests the robust of using identical $\delta V_{\text{HL}}$, $\Delta_{\text{HL}}$, and $\delta$ to depict the variants within a whole QD class.

\paragraph*{The origin of supercoupling between HH and LH.}
As discussed above,  there is a dense manifold of HH-like intermediate states lying between HH0 and LH0 in strained In(Ga)As/GaAs QDs, and a schematic neglected case is shown in Fig.~\ref{fig:schematic}(b). In $C_{2v}$ symmetry  QDs, such HH-like intermediate states belong to same symmetry representation $\Gamma_5$ as HH0 and LH0. Therefore, the perturbation potential  $\delta V_{C_{2v}}$  induces the inter-band coupling between HH and LH states, but also coupling among HH states which was unrecognised in literature to describe the $h_0$ state. The coupling among HH-like QD states is subject to additional terms:
 \begin{equation}
 \label{eq4-3}
|\Psi_{h_0}\rangle = |\Psi'_{\textrm{HH}0}\rangle + 
\sum_{n\geqslant 1}\frac{\langle \Psi'_{\textrm{HH}n}|\delta V_{C_{2v}}|\Psi'_{\textrm{HH}0}\rangle}{E'_{\textrm{HH}0}-E'_{\textrm{HH}n}}|\Psi'_{\textrm{HH}n}\rangle,
\end{equation}
where $|\Psi'_{\textrm{HH0}}\rangle$ is defined in Eq.~(\ref{eq4-2}) and $E'_{\textrm{HH}n}$ is the energy of HH-like excited states $|\Psi'_{\textrm{HH}n}\rangle$ ($n=1,2, ...$).  After inserting Eq.~(\ref{eq4-2}) into Eq.~(\ref{eq4-3}), we are ready to obtain revised $\lambda_{\textrm{LH0}}$,
 \begin{equation}
 \label{eq4-4}
\lambda_{\textrm{LH0}} = \frac{\langle \Psi^0_{\textrm{LH0}}| \delta V_{C_{2v}}|\Psi^0_{\textrm{HH0}}\rangle}{E^0_{\textrm{HH0}}-E^0_{\textrm{LH0}}} + \sum_{n\geqslant1}\frac{\langle \Psi^0_{\textrm{HH}n}| \delta V_{C_{2v}}|\Psi^0_{\textrm{HH0}}\rangle}{E'_{\textrm{HH0}}-E'_{\textrm{HH}n}}\cdot \frac{\langle \Psi^0_{\textrm{LH0}}| \delta V_{C_{2v}}|\Psi^0_{\textrm{HH}n}\rangle}{E^0_{\textrm{HH}n}-E^0_{\textrm{LH0}}}  + \mathcal{O}(E^{-3}).
\end{equation}
The additional terms mediated by HH-like excited states $\text{HH}n$ could be regarded as higher order terms of a Taylor  series  of
\begin{equation}
\label{eq4-5}
\lambda_{\textrm{LH0}} = \frac{\langle \Psi^0_{\textrm{LH0}}| \delta V_{C_{2v}}|\Psi^0_{\textrm{HH0}}\rangle}{E^0_{\textrm{HH0}}-E^0_{\textrm{LH0}}-\delta}.
\end{equation}
The parameter $\delta$ is adjustable to accommodate the difference between Eq.~(\ref{eq4-4}) and Taylor  series. It is now manifested that the effective reduction of HH-LH splitting $\Delta_{\text{HL}}$  by $\delta$ originates from indirect coupling between HH0 and LH0 mediated  by HH-like excited states, in  analogy to well known superexchange magnetic interaction through a non-magnetic anion \cite{Anderson50} . We refer this novel indirect coupling channel as supercoupling. Because of  large HH-LH splitting in strained QDs, the supercoupling effect will dominant the HH-LH mixing over  the direct coupling between HH0 and LH0.  Specifically, if it is absence of supercoupling, say $\delta=0$, the magnitude of the HH-LH mixing, $\lambda^2_{\textrm{LH}}$, will tend to less than 1\%  instead of  5-20\% as predicted in strained In(Ga)As/GaAs QDs, even though the strain significantly enhances the coupling matrix $\delta V_{\text{HL}}$ by a factor of 4.5 with respect to unstrained GaAs/AlGaAs QDs.  The supercoupling of HH and LH is further confirmed by an abnormal  point within  the class of GaAs QDs  (indicated by \#1 in Fig.~\ref{fig:Mixing}). For this specific QD the coupling between $h_0\approx| \Psi^0_{\textrm{HH}}\rangle$ and $h_2\approx| \Psi^0_{\textrm{LH}}\rangle$ is mediated by a HH-dominated QD state ($h_1$), whereas in the remaining QDs of its family the state  $h_0\approx| \Psi^0_{\textrm{HH}}\rangle$ is immediately followed by $h_1\approx| \Psi^0_{\textrm{LH}}\rangle$.

\begin{figure}[H]
\centering
\includegraphics[width=0.8\textwidth]{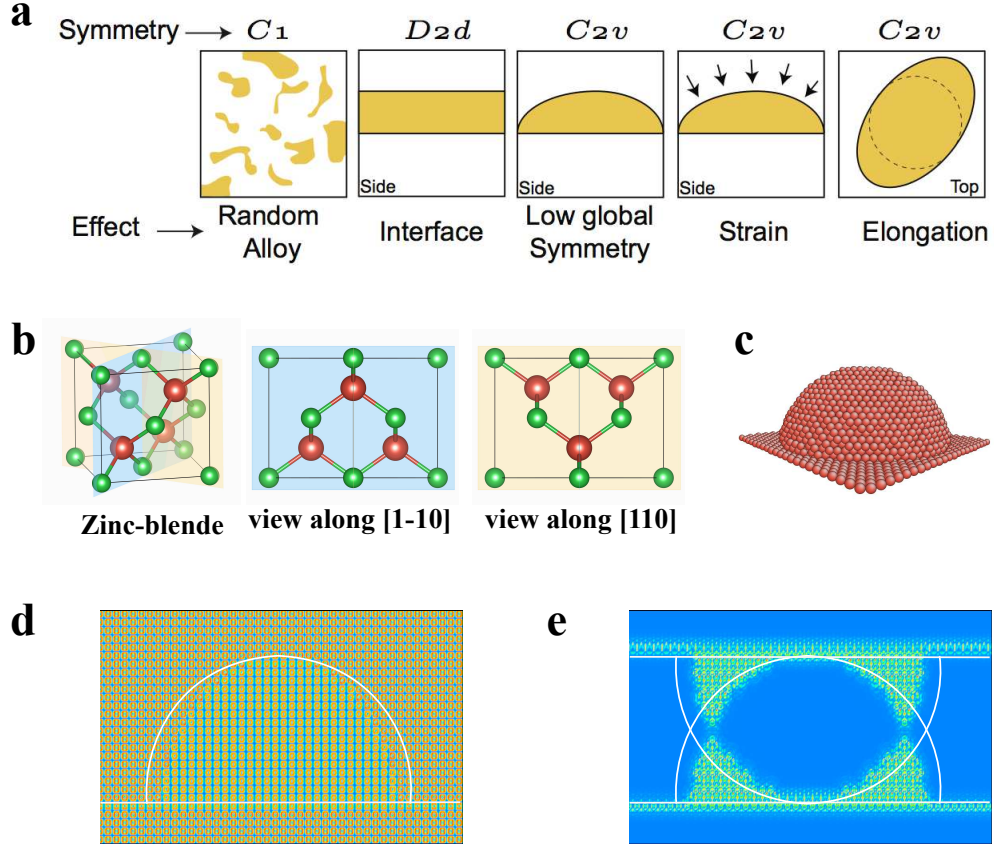}
\caption{\label{fig:Effects} {\bf Illustrations of the effects leading to non-vanishing HH-LH coupling.} {\bf a.} Schematic illustration of the six possible physical effects leading to non-vanishing direct HH-LH coupling matrix element $\delta V_{\text{HL}}$. {\bf b.} Zinc-blende crystal structure with views along the [1$\bar{1}$0] and [110] directions. The yellow and blue planes represent the crystal (110) and (1$\bar1$0) planes. {\bf c.} Schematic of a lens-shaped GaAs QD sitting on a one monolayer thick GaAs wetting layer. {\bf d.} Absolute difference of the crystal potential, of the QD defined in c, in the (110) plane and in the (1$\bar 1$0) plane. Red for maximum and blue for zero value.  The expected potential difference is zero if the QD has $D_{2d}$ symmetry as supposed in continuum theory.  {\bf e.} Absolute difference of the crystal potential in the (110) plane and the potential in the (1$\bar 1$0) plane but after a reflection operation about the (001) mirror plane.  The areas of finite values near the interfaces of the QD show therefore deviations from $D_{2d}$ symmetry.}
\end{figure}

\paragraph*{Physical factors affecting the direct HH-LH coupling matrix element $\delta V_{\text{HL}}$. } Having discussed the denominator effect $\Delta_{\text{HL}}$ in   Eq.~(\ref{eq:super}) quantifying the enhancement of HH-LH mixing through the  supercoupling mechanism, we next discuss the numerator effects $\delta V_{\text{HL}}$   quantifying  the  relative  importance  of   distinct  factors  leading  to  HH-LH mixing. An important observation here is the role of atomically resolved symmetry {\it vs} the global shape symmetry. As shown in Figs.~\ref{fig:Effects}b-e, the  symmetry  of  ideally  shaped  (circular based  lens-, cone-, and Gaussian-shape) QDs made of common zincblende structure semiconductors is already $C_{2v}$ as evidenced by the  inequivalent [110] and [1$\bar1$0] directions illustrated in Fig.~\ref{fig:Effects}d. QDs are distinct from bulk zinc-blende crystals or $D_{2d}$ symmetric (001) quantum wells, where the [110] direction can be transformed to the [1$\bar1$0] direction by an $S_4$ symmetry operations (90$^\circ$ rotation followed by a reflection  \cite{Koster63}). Thus, whereas in the latter systems symmetry forbids HH-LH mixing, in ideally shaped  QDs embedded in a matrix  {\it the HH-LH  mixing  is  intrinsically  allowed}   even  without  built-in  strain  or  QD  shape non-ideality (anisotropy).  The mechanisms that contribute to the direct HH-LH coupling matrix element $\delta V_{\text{HL}}$ are analyzed and quantified next.

\paragraph*{(1) 3D confinement of wave functions in QDs has but a negligible effect  on $\delta V_{\text{HL}}$.} In the  Luttinger-Kohn Hamiltonian (or other  {$k \cdot p$}  methods) the HH-LH mixing is  present even in 3D bulk at nonzero wavevector $\bf k$, because of the finite off-diagonal band coupling terms  $R$ and $S$ (see the method section), and is absent at zone-center ${\bf k}=0$ where $R$ and $S$ vanish. In 0D QDs the electronic states have a finite effective wavevector as a result of 3D confinement, since ${\bf k}_i$ is replaced by the operator $i \frac{\partial} {\partial r_i}$ ($i=x,y,z$), leading to finite $R$ and $S$. Therefore, in the Luttinger-Kohn formalism applied to nanostructures the HH-LH mixing is formally always present, even in cylindrically symmetric QDs \cite{SSLi96, fischer10,Tanaka93,liao12, LegerPRB07}.  Such 3D quantum confinement  was previously considered as the only mechanism leading to HH-LH mixing in unstrained QDs  \cite{Tanaka93, fischer10, liao12}.  In an unstrained and flat [where $a_z \text{(height)}\ll L \text{(wide)}$] GaAs QDs, this coupling within the Luttinger-Kohn Hamiltonian gives rise to the dependence $\lambda_{\textrm{LH}}\simeq 0.53 a_z/L$ \cite{fischer10}. This relationship predicts  $\lambda^2_{\text{LH}}=0.2$\% which is significantly lower than our determined 3.5\%  for a disk-shaped GaAs QD ($a_z=2$ nm, $L=25.2$ nm). Furthermore, Fig.~\ref{fig:Mixing}b shows that in an atomistic calculation $\lambda^2_{\textrm{LH}}$ is nearly insensitive to the QD height for both disk-shaped  and  lens-shaped  GaAs/Al(Ga)As flat QDs; in contradiction to the model Hamiltonian result of  $\lambda^2_{\textrm{LH}}\simeq (a_z/L)^2$. These results demonstrate the negligible effect of 3D confinement on HH-LH mixing in flat QDs. We should note that the fact that the atomistic symmetry ultimately controls the existence of HH-LH coupling was usually overlooked in {$k \cdot p$} calculations following the above description. Even when $R$ and $S$ terms are finite, the HH-LH mixing should be absent if such mixing is forbidden by symmetry. The well-known HH-LH mixing away from the $\Gamma$-point in 3D bulk is not only due to finite $R$ and $S$, but also due to the reduced symmetry of these k-points.

\paragraph*{(2) Shape  anisotropy (e.g in-plane elongation) in QDs has  but a  small  effect  on $\delta V_{\text{HL}}$:} The in-plane shape anisotropy (Fig.~\ref{fig:Effects}a), can lower the QD symmetry from $D_{2d}$  to $C_{2v}$ and enance the HH-LH mixing. In the Luttinger-Kohn Hamiltonian its contribution to HH-LH mixing is described by the $R$ term associated with $k^2_x-k^2_y$  and is $a_z/L$ times smaller than the $S$ term associated with $k_z$ \cite{Bir-Pikus}. The directly calculated $\delta V_{\text{HL}}$  values of Fig.~\ref{fig:Mixing}a include seven GaAs QDs with an anisotropic shape (6 QDs are elongated along the [1$\bar1$0] direction and one along the [110] direction) and nominal  $C_{2v}$ symmetry.   Interestingly, however, the HH-LH mixing magnitude $\lambda^2_{\textrm{LH}}$ of  these   irregularly shaped QDs fall on  the same  universal  curve  as the 17 circular based QDs (as shown in Fig.~\ref{fig:Mixing}) indicating  the  minor  role of  shape  anisotropy  on  HH-LH  mixing.    This finding highlights the incorrect link often drawn between HH-LH mixing and shape anisotropy, whereby one infers a shape anisotropy from the measured coupling \cite{belhadj10,Moehl06,JFR06,Testelin09}.

\paragraph*{(3) Build-in strain does not lower the symmetry but enhances $\delta V_{\text{HL}}$.} 
In the classical  Pikus-Bir strain Hamiltonian \cite{wei94, Bir-Pikus}, the shear strain components ($\epsilon_{xy}$, $\epsilon_{yz}$ and $\epsilon_{zx}$), belonging to rhombohedral symmetries, give rise to finite off-diagonal $R$ and $S$ terms  (see the method section), which will mix HH and LH if such mixing is allowed by symmetry. These shear components are absent in bulk $D_{2d}$ and $C_{2v}$ but are present at the interfaces of QDs. From an atomistic point of view, the built-in strain does not lower the symmetry and is, as such, not the reason for the creation  of  HH-LH coupling. However, such built-in strain  through the atomic relaxation allows the local asymmetry of the interface to propagate inside the QD, where the wave functions are localized \cite{bester05a}. The increase in $\delta V_{\text{HL}}$ (2.15 to 9.82 meV), from blue triangles of unstrained GaAs/Al(Ga)As QDs to green dots of strained In(Ga)As/GaAs QDs as shown in Fig.~\ref{fig:Mixing}a, is mainly due to the built-in strain. We conclude that the build-in strain constitutes an important contribution to the coupling [$(9.82-2.15 )/9.82=78 \%$ in strained QDs]. 

\paragraph*{(4) Alloy  disorder in the QD material or its matrix has but a small effect on $\delta V_{\text{HL}}$:}  Although the alloy randomness is important for both exciton fine structure splitting \cite{luo12} and optical polarization \cite{mlinar09} in QDs, it has a negligible effect on HH-LH mixing  as demonstrated here by the fact that  both ordered InAs/GaAs and disordered In$_{60}$Ga$_{40}$As/GaAs QDs share the same $\delta V_{\text{HL}}=9.82$ meV and both ordered GaAs/AlAs and disordered GaAs/Al$_{30}$Ga$_{70}$As QDs share the same $\delta V_{\text{HL}}=2.15$ meV. Moreover, five different random alloy realizations of a Gaussian-shaped  3 nm heigh GaAs/Al$_{30}$Ga$_{70}$As QD lead to virtually the same coupling $\lambda^2_{\textrm{LH}}= 13.3, 13.0, 12.9, 13.1, 12.8\%$, with a standard deviation of $\sigma=0.2\%$. Also, five different random alloy realizations of In$_{60}$Ga$_{40}$As/GaAs  QDs  give rise to similar four $\lambda^2_{\textrm{LH}} = 3.6\%$ and one $\lambda^2_{\textrm{LH}}=3.8\%$. This strongly suggests negligible alloy disorder effect on HH-LH mixing.  
  
\paragraph*{(5) Significant effect of low-symmetry interfaces on $\delta V_{\text{HL}}$:}  (001)-quantum  wells BAB with  a  global $D_{2d}$  symmetry  consist of two  interfaces B-A and A-B each having  a lower,  $C_{2v}$ local  symmetry. When   both $C_{2v}$ interfaces  are  considered simultaneously the  mirror  plane  operation in the well center
joins them into the higher $D_{2d}$  point  group. A periodic bulk material with this symmetry has zero HH-LH mixing at the $\Gamma$-point. The quantum well geometry introduces folding along the well direction and formally allows HH-LH mixing at the $\bar{\Gamma}$-point represented by a finite value of $S$ in the envelope function approach. However, this mixing effect is very small and significantly underestimates the full mixing, as we have described in point {\it (1)}. It was recognized long ago \cite{ivchenko96} that the local $C_{2v}$ symmetry of each individual interface in a quantum well, ignored in standard $k \cdot p$ approaches, gives rise to the HH-LH mixing. We have considered the analogous situation in nominal $D_{2d}$ (disk-shaped) GaAs QDs embedded in AlAs and Al$_{30}$Ga$_{70}$As barriers. In Fig.~\ref{fig:Mixing} we also show $\lambda^2_{\textrm{LH}}$ for such GaAs QDs (represented by filled magenta triangles). We see that the HH-LH mixing due  to the  local  $C_{2v}$  interface  effect is $\delta V_{\text{HL}}=0.8$ meV in both cases. This represents around  $0.8/2.15 \sim  40\%$ of the total coupling strength $\delta V_{\textrm{HL}}$ in $C_{2v}$ unstrained GaAs QDs and  $0.8/9.82\sim 8\%$ in strained InAs QDs. 

\paragraph*{(6) Effect of the intrinsic $C_{2v}$ symmetry of symmetric QDs on $\delta V_{\text{HL}}$ due to the inequivalence of the [110] and [1$\bar1$0] directions.} The  atomistic symmetry of ideal circular based  lens-shape or Gaussian-shape QDs is $C_{2v}$, and therefore lower than in the previously discussed quantum well case,  because the [110] and [1$\bar1$0] directions are nonequivalent due to the upper curved interface, which is illustrated in Figs.~\ref{fig:Effects}d and c.  The curvature introduced in the top interface corresponds to a lowering of the global symmetry;  from a disk-shaped QD with global $D_{2d}$ symmetry to a lens-shaped QD with a global $C_{2v}$ symmetry. The increase in $\delta V_{\text{HL}}$ from disk-shaped QDs  $\delta V_{\text{HL}}=0.8$ meV  to lens-shaped QDs  $\delta V_{\text{HL}}=2.15$ meV  (red to blue triangles) is attributed to the intrinsic $C_{2v}$ symmetry due to the bending of the top interface, i.e. due to an asymmetry in the growth direction. We therefore conclude that the intrinsic $C_{2v}$ symmetry of symmetric QDs  is responsible for 60\% of $\delta V_{\text{HL}}$  in unstrained  $C_{2v}$ GaAs/AlGaAs QDs and for 14\% in strained InGaAs/GaAs QDs.

\paragraph*{Discussion.} For unstrained self-assembled QDs effects (1)-(6) contribute 0\%, 0\%, 0\%, 0\%, 40\%, and 60\%, whereas for strained self-assembled QDs they contribute 0\%, 0\%, 78\%, 0\%, 8\%, and 14\% to the direct HH-LH coupling matrix element $\delta V_{\text{HL}}$, which is a numerator of HH-LH mixing $\lambda^2_{\textrm{LH}}$. However, the supercoupling effect, via dense intermediate QD states, is the dominant mechanism for finite HH-LH mixing in strained QDs. Only the reduction of the denominator (Eq.\ref{eq:super}) as a consequence of supercoupling can lead to the HH-LH mixing $\lambda^2_{\textrm{LH}}$ values we obtain from our atomistic calculations. 

With the new understanding of HH-LH mixing presented in this paper, we are likely to propose design rules to optimize HH-LH mixing in QDs to the values required by specific opto-electronic and spintronic applications. If one needs QDs with weaker HH-LH mixing and a narrow variation within the sample, it is best to choose unstrained $D_{2d}$ QDs. But they are fairly rare in experimental synthesis.  For usual self-assembled QDs,  the direct HH-LH matrix element $\delta V_{\textrm{HL}}$ is relatively insensitive to the QD morphology for a class of QDs, however the magnitude of HH-LH mixing $\lambda^2_{\textrm{LH}}$ could be effectively tuned by engineering the effective HH-LH splitting $\Delta_{\text{eff}}$.   Specifically, in the class of {\it strained} In(Ga)As/GaAs QDs, where the finite HH-LH mixing is mainly due to the supercoupling of HH and LH meditated by intermediate QD states, flatter  QDs  have larger $\Delta_{\text{eff}}$ and  consequently smaller HH-LH mixing   $\lambda^2_{\textrm{LH}}$. In the class of {\it unstrained} self-assembled GaAs/Al(Ga)As QDs, the QD-height does not control the HH-LH mixing as a result of the independence of its two main mechanisms  on QD-height, but the lens-shape  often exhibit weaker HH-LH mixing than  than Gaussian-shape. Increasing the Al composition of the barrier for GaAs/Al(Ga)As QDs leads to enhance HH-LH splitting and to reduced HH-LH mixing.

{\bf Methods} 
\paragraph*{Pseudopotential calculations.} 
The electronic states of  GaAs/Al(Ga)As and In(Ga)As/GaAs QDs
are obtained by solving the Schr\"{o}dinger equation in a  crystal (QD+matrix) potential $V(\bf r)$ 
within a basis of strained bulk Bloch bands \cite{wang99b}. The screened potential $V(\bf r)$ is  described as a superposition of atomic 
pseudopotentials $\hat{v}_{\alpha}$ centered at the atomic positions ${\bf R}_{\alpha,n}$  \cite{wang99b},  where $n$ is the primary cell site index:  $V({\bf r})=\sum_n\sum_{\alpha}\hat{v}_{\alpha}({\bf r}-{\bf R}_{\alpha,n})$.  This approach captures the multiband, intervalley and spin-orbit interactions and also forces upon the eigenstates the correct atomistic symmetry of the underlying nanostructure. 
The atomic pseudopotentials $\hat{v}_{\alpha}$ are fit to experimental
transition energies, effective masses, spin-orbit splittings and deformation potentials of the underlying bulk semiconductors as well as to band offets of e.g., InAs/GaAs, heterjunctions \cite{williamson00, luo09}.
The  atomistic valence force field (VFF) model \cite{Pryor98} is used to find the equilibrium atomic positions ${\bf R}_{\alpha,n}$ via minimization of the lattice-mismatch induced strain energy. 
A real space Pikus-Bir Hamiltonian \cite{wei94}  is used to extract strain-modified confinement potentials in QDs \cite{Pryor98}. Fig.~\ref{fig:BandOffsets} shows strain-modified electron and hole confinement potentials of an  InAs/GaAs QD in comparison to strain-free potentials of a GaAs/AlGaAs QD.  

\paragraph*{Luttinger-Kohn and Pikus-Bir Hamiltonians.} 
According to Pikus and Bir \cite{Bir-Pikus}, the correspondence between the strain Hamiltonian and the Luttinger-Kohn Hamiltonian is 
\begin{equation}
\label{strain}
k_ik_j \leftrightarrow e_{ij},
\end{equation}
therefore the total Hamiltonian $H=H_{\bf k\cdot p}+H_{strain}$ describing the top of the valence band for bulk zinc-blende or diamond semiconductors under strain $e$ is given by
\begin{equation}
\label{LK}
H=\begin{pmatrix}
      P+Q &-S & R & 0    \\
      -S^{\dag}&  P-Q & 0 & R \\
      R^{\dag} &0 & P-Q & S \\
      0 & R^{\dag} &S^{\dag} & P+Q
\end{pmatrix},
\end{equation}
where the spin-orbit split-off band is ignored and all matrix elements are written in terms of three dimensionless Luttinger parameters $\gamma_1$, $\gamma_2$, and $\gamma_3$ and three deformation potentials $a$, $b$, and $d$:
\begin{equation}
P=\frac{\hbar^2}{2m}\gamma_1 k^2-a_v(e_{xx}+e_{yy}+e_{zz}),
\end{equation}
\begin{equation}
Q=\frac{\hbar^2}{2m}\gamma_2 (k_x^2+k_y^2-2k_z^2)-\frac{b}{2}(e_{xx}+e_{yy}-2e_{zz}),
\end{equation}
\begin{equation}
S=\frac{\hbar^2}{2m}2\sqrt{3}\gamma_3 (k_x-ik_y)k_z-d(e_{xz}-ie_{yz}),
\end{equation}
\begin{equation}
R=\frac{\hbar^2}{2m}\sqrt{3}[-\gamma_2(k_x^2-k_y^2)+2i\gamma_3k_xk_y]+\frac{\sqrt{3}}{2}b(e_{xx}-e_{yy})-ide_{xy}.
\end{equation}
The basis vectors are the four degenerate Bloch wave functions (HH and LH bands) at the center of the Brillouin zone:
\begin{equation}
|3/2,3/2\rangle=-\frac{1}{\sqrt{2}}|(X+iY)\uparrow \rangle,
\end{equation}
\begin{equation}
|3/2,1/2\rangle=\frac{1}{\sqrt{6}}|-(X+iY)\downarrow+2Z\uparrow \rangle,
\end{equation}
\begin{equation}
|3/2,-1/2\rangle=\frac{1}{\sqrt{6}}|(X-iY)\uparrow+2Z\downarrow \rangle,
\end{equation}
and
\begin{equation}
|3/2,-3/2\rangle=\frac{1}{\sqrt{2}}|(X-iY)\downarrow \rangle.
\end{equation}

\section*{ACKNOWLEDGMENTS}
JL was supported by the National Young 1000 Talents Plan and the National Science Foundation of China (NSFC grant \#61474116). AZ was supported by Office of Science, Basic Energy Science, MSE division under grant DE-FG02-13ER46959 to CU Boulder. GB was supported by the BMBF (QuaHL-Rep, Contract No. 01BQ1034).


\FloatBarrier
\newpage

%

\FloatBarrier
\newpage
\bibliographystyle{plainnat}
\bibliography{qwe}

\begin{thebibliography}{41}
\bibitem{note1}{Sentences such as ``... the existence of heavy and light hole mixing in our QDs caused by the QD in-plane anisotropy" \cite{xu07} can be found in many places of the QD literature.}.
\bibitem{note2}
The morphology of SAQDs calculated by atomistic methods  are input from experimental measurements. The QD height varies from 2 to 6 nm. In the case of strained In(Ga)As/GaAs QDs the base size is in the rang from 20 nm to 26 nm and aspect ratio of QD base is from 0.8 to 1.2.  strain-free GaAs/Al(Ga)As and strained In$_{0.6}$Ga$_{0.4}$As/GaAs QDs. The Ga composition in the AlGaAs alloy barrier is 70\%.  In the case of unstrained GaAs/AlGaAs QDs, the base size is 30 nm and 35x30 nm for circular  and elongated Gaussian-shaped QDs, respectively.
\end{thebibliography}

\end{document}